\documentclass[11pt]{article}
\usepackage{amssymb,latexsym, amsmath}
\usepackage{graphicx}
\usepackage{epsfig}
\newcommand{\remfigure}[1]{#1}

\newcommand{\comment}[1]{\vspace{-3 mm}\par
\marginpar{\large\underline{}}\noindent
\framebox{\begin{minipage}[c]{0.65 \textwidth}
\rm #1 \end{minipage}}\vspace{1 mm}\par}
\newcommand{\rem}[1]{}
\def\mathbf#1{{\bf #1}}
\def\mathsf#1{{\sf #1}}
\def\mathbb#1{\mathbf{ #1}}
\def\mathfrak#1{\mathbf{ #1}}

\begin{document}

\title{
\vbox to 0pt {\vskip -1cm \rlap{\hbox to \textwidth {\rm{\small
BI3692
\hfil
ACCEPTED FOR {\it PHYS. LETT. A}
\hfil
%Today's version $\quad$
\today}}}}
Nonlinear balance and exchange of stability in dynamics of
solitons, peakons, ramps/cliffs and leftons in a
1+1 nonlinear evolutionary pde
%\vspace{-2mm}
}
\author{
Darryl D. Holm$^{1,2}$ and Martin F. Staley$^1$
\\$^1$Theoretical Division and Center for Nonlinear Studies
\\Los Alamos National
Laboratory, MS B284
\\ Los Alamos, NM 87545
\\ email: dholm@lanl.gov, mstaley@lanl.gov
\and
$^2$%
Mathematics Department\\
Imperial College of Science, Technology and Medicine
\\ London SW7 2AZ, UK
}

\date{
\comment{
\centerline{BI3692}
\centerline{ACCEPTED FOR {\it Phys. Lett. A}}
\centerline{January 16, 2003}
}
}

\maketitle

%\large

%\vspace{-5mm}

\begin{abstract}
%%%%%%%%%%%%%%%%%%%%%%%%%%%%%%%%%%%%%%%%%%%%%%%%%%%%%%%%%%%%%%%%%%%%%%%%
We study exchange of stability in the dynamics
of solitary wave solutions
under changes in the nonlinear balance in
a  1+1 evolutionary partial differential equation
related both to shallow water waves and to turbulence.
We find that solutions of the equation
$
m_t
+
um_x
+b\,u_xm
=
\nu\,m_{xx}
$
with $m = u - \alpha^2 u_{xx}$ for fluid velocity $u(x,t)$ change their
behavior at the special values $b=0,\pm1,\pm2,\pm3$.
\\
%%%%%%%%%%%%%%%%%%%%%%%%%%%%%%%%%%%%%%%%%%%%%%%%%%%%%%%%%%%%%%%%%%%%%%%%



\end{abstract}

\newpage
\tableofcontents

\vskip 4mm

%\begin{multicols}{2}

\section{Introduction}
%%%%%%%%%%%%%%%%%%%%%%%%%%%%%%%%%%%%%%%%%%%%%%%%%%%%%%%%%%%%%%%%%%%%%%%%
\subsection{Linear and nonlinear balances in shallow water waves}
%%%%%%%%%%%%%%%%%%%%%%%%%%%%%%%%%%%%%%%%%%%%%%%%%%%%%%%%%%%%%%%%%%%%%%%%
The primary physical mechanism for the propagation of solitary shallow water
waves is the balance between nonlinear steepening and linear dispersion.
This balance appears {\it uniquely} at linear order in an asymptotic
expansion in the Korteweg-de Vries equation (KdV),
\begin{equation}\label{kdv}
u_t + c_0u_x + \frac{3\epsilon_1}{2}uu_x +
\frac{3\epsilon_2}{20}\,u_{xxx}=0
\,.
\end{equation}
Here the expansion parameters satisfy $\epsilon_1 \ge \epsilon_2 >
\epsilon_1^2$ and are defined by $\epsilon_1 = a/h$ and $\epsilon_2 =
h^2/l^2$, in terms of $a$, $h$ and $l$, which denote wave amplitude, mean
water depth, and typical horizontal length scale (e.g., a wavelength),
respectively. KdV possesses the famous
${\rm sech}^2$ solitary wave solution $u(x,t)=u_0\,{\rm
sech}^2((x-ct)\sqrt{u_0/\gamma}/2\,)$, for
$u_0=2(c-c_0)/\epsilon_1$ and $\gamma=3\epsilon_2/(5 \epsilon_1)$, see
\cite{AblowitzClarkson[1991]}.

The Benjamin-Bona-Mahoney equation (BBM),
\begin{equation}\label{bbm}
u_t + c_0u_x + \frac{3\epsilon_1}{2}uu_x
- \frac{3\epsilon_2}{20} c_0^{-1}\,u_{xxt}
=
0
\,,
\end{equation}
has the same sech$^2$ shape for its solitary wave, but with $\gamma$
replaced by $\gamma\,'=c\gamma/c_0$. BBM is asymptotically equivalent to KdV
at order $O(\epsilon_1,\epsilon_2)$. However, the linear dispersion
relation for BBM is a better match than that for KdV in the physical
application to shallow water waves.

Beyond KdV at linear order, the asymptotic expansion at quadratic order
in the small parameters $\epsilon_1$ and $\epsilon_2$ does not
produce a unique wave equation for unidirectional shallow water
waves \cite{DGH[2003]}. Instead, the asymptotic expansion
produces an entire {\it family} of shallow water wave equations that are
asymptotically equivalent to each other at quadratic order in the shallow
water expansion parameters. The equations in this family are related
amongst themselves by a continuous, three-parameter group
of nonlinear, nonlocal transformations of variables given by,
\begin{equation}
   \label{KodamaKdV}
     u =
v + \epsilon_1(a_1v^2+a_2v_x\partial^{-1}_x{v})
   + \epsilon_2a_3 v_{xx}
\,,
\end{equation}
in which $(a_1,a_2,a_3)$ are real parameters. This transformation group
was first introduced for determining normal forms of shallow water
equations by Kodama in
\cite{Kodama[1985&1987]}.

Among the family of asymptotically equivalent shallow water wave
equations at quadratic order accuracy in the small parameters $\epsilon_1 =
a/h$ and $\epsilon_2 = h^2/l^2$ are several equations that are completely
integrable. As for KdV at linear order, these integrable shallow water
equations at  quadratic order possess soliton solutions that interact via
elastic collisions.  In particular, the equation in the KdV hierarchy with
fifth-order derivatives is present among these integrable equations, as
shown in \cite{LiZhiSibgatullin[1997]}.

The family of shallow water equations at quadratic order accuracy  that are
asymptotically equivalent under Kodama transformations (\ref{KodamaKdV})
contains the following sub-family, derived in \cite{DGH[2003]},
\begin{equation}
\label{b-eqn}
\underbrace{\
m_t \
}_{\hbox{Evolution}}
\!\!\!+
\underbrace{\
c_0u_x
+ \frac{3\epsilon_2}{20}u_{xxx} \
}_{\hbox{Dispersion}}
+
\underbrace{\
\epsilon_1(um_x + b\,mu_x) \
}_{\hbox{Nonlinearity}}
= 0\,,
\end{equation}
where $m = u - (19\epsilon_2/60) u_{xx}$. The set of 1+1 evolutionary
equations (\ref{b-eqn}) is nonlocal, dispersive and nonlinear. The
nonlinearity in this set of asymptotically equivalent shallow water
equations is parameterized by the real constant $b$, which depends on the
group parameters $(a_1,a_2,a_3)$ in the Kodama transformation
(\ref{KodamaKdV}).  An asymptotically equivalent shallow water equation for
any $b \not = -1$ may be achieved by a Kodama transformation. However, the
case $b=-1$ violates the asymptotic ordering and the corresponding Kodama
transformation is singular in this case \cite{DGH[2003]}.

The cases $b=2$ and $b=3$ are special values for the $b-$equation
(\ref{b-eqn}). The case $b=2$  restricts (\ref{b-eqn}) to
the integrable Camassa-Holm equation (CH) \cite{CH[1993]}.
The case $b=3$ in (\ref{b-eqn}) is the Degasperis-Procesi equation (DP)
\cite{DP[1999]}, which was shown to be integrable in \cite{DHH2002}.
These two cases exhaust the integrable candidates for (\ref{b-eqn}), as
was shown using Painlev\'e analysis in \cite{DHH2002}. The b-family of
equations (\ref{b-eqn}) was also shown in \cite{MN[2002]} to admit the
symmetry conditions necessary for integrability only in the cases $b=2$ for
CH and
$b=3$ for DP.

KdV (\ref{kdv}) and the cases $b=2$ (CH) and $b=3$ (DP) of equation
(\ref{b-eqn}) are three completely integrable Hamiltonian equations
that possess solitons as traveling waves.  In KdV, the balance that
confines the traveling wave soliton occurs between nonlinear
steepening and linear dispersion. This is the leading order asymptotic
balance for shallow water waves. However, even in the absence of linear
dispersion, the parameter $b$ in equation (\ref{b-eqn}) introduces
additional possibilities for balance, including the nonlinear/nonlocal
balance in the following (rescaled) dispersionless  case of CH that was
studied previously for
$b=2$ in  \cite{CH[1993]},
\begin{equation}
\label{b-family-inviscid}
m_t + um_x + b\,mu_x = 0\,,
\hbox{ with }
m = u - \alpha^2 u_{xx}
\,.
\end{equation}
The nonlinear/nonlocal balance in this equation, even in the absence of
linear dispersion, can still produce a confined solitary traveling wave
pulse $u(x,t)=ce^{-|x-ct|/\alpha}$, called the {\it  peakon}
\cite{CH[1993]}. The peakon moves with speed equal to its amplitude and has
a jump in derivative at its peak. Peakons for either $b=2$ or $b=3$ are
true solitons that interact via elastic collisions under CH dynamics, or DP
dynamics, respectively. In addition, the CH and DP initial value problems
are both completely integrable as Hamiltonian systems by using the inverse
spectral transform (IST) method for an isospectral linear eigenvalue
problem whose purely discrete spectrum gives the asymptotic speeds of the
peakons
\cite{CH[1993]}, \cite{DHH2002}. Figure \ref{Gaussian-to-Peakons-CH} shows
the evolution under dispersionless CH for the case $b=2$ in
equation (\ref{b-family-inviscid} ) of a Gaussian initial
velocity distribution of unit area and width $5\alpha$. Peakon
solutions exist for equation (\ref{b-family-inviscid}) with any value of
$b$. However, we shall find that the stability of these peakon solutions
requires
$b>1$.

The properties of dispersionless CH for the case $b=2$ in
equation (\ref{b-family-inviscid}) and also the  related class of
dispersionless equations consisting of (\ref{b-family-inviscid}) with
$u=g*m=\int_{-\infty}^\infty g(x-y)m(y)dy$ for the case $b=2$ and an even
kernel $g(x)=g(-x)$ including $g(x)=e^{-|x|/\alpha}$ are studied further in
Fringer and Holm \cite{FH[2001]}. The properties of dispersionless DP for
the case $b=3$ in equation (\ref{b-family-inviscid}) are studied further in
\cite{DHH2002}, \cite{HS2003}.

%\vspace{6.5cm}

%%%%%%%%%%%%%%%%%%%%%%%%%%%%%%%%%%%%%%%%%%%%%%%%%%%%%%%%%%%%%%%%%%%%%%%%
\rem{
\begin{figure}
\begin{center}
   \leavevmode{\hbox{\epsfig{
       figure=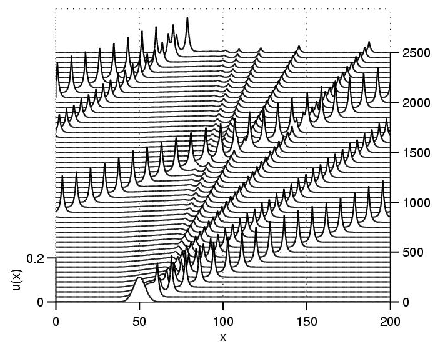, scale=0.7
    }}
}
\caption{\label{Gaussian-to-Peakons-CH}
    Evolution under equation (\ref{b-family-inviscid}) of a Gaussian initial
velocity
    distribution of unit area and width $5\alpha$.
}
\end{center}
\end{figure}
}

%\vskip-6.5cm
\remfigure{
\begin{figure}[htbp]
\centering
\psfig{file=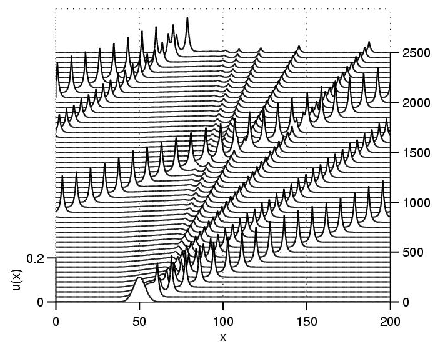,height=6.5cm}
\caption[]{
Evolution under equation (\ref{b-family-inviscid}) with $b=2$ of a
Gaussian initial velocity distribution of unit area and width $5\alpha$.
}
\label{Gaussian-to-Peakons-CH}
\end{figure}
}
%%%%%%%%%%%%%%%%%%%%%%%%%%%%%%%%%%%%%%%%%%%%%%%%%%%%%%%%%%%%%%%%%%%%%%%%

The dispersionless limit of KdV in (\ref{kdv}), upon rescaling velocity
$u$, yields the Burgers equation
\begin{equation}
u_t + uu_x - \nu\,u_{xx}=0
\,,
\end{equation}
in which we now also add constant viscosity $\nu$. The characteristic
Burgers solution is the classic ramp and cliff shown arising from the
Gaussian initial condition in Figure
\ref{Gaussian-to-ramp/cliff-Burgers}. In the ramp/cliff solution, nonlinear
steepening is balanced by linear viscosity to produce the ``cliff'' whose
width is controlled by the magnitude of viscosity $\nu$. The ``ramp'' is
the self-similar $u\approx{x}/t$ part of the solution for which the viscous
term vanishes.

%\vspace{6.5cm}

%%%%%%%%%%%%%%%%%%%%%%%%%%%%%%%%%%%%%%%%%%%%%%%%%%%%%%%%%%%%%%%%%%%%%%%%
\rem{
\begin{figure}
\begin{center}
   \leavevmode{\hbox{\epsfig{
       figure=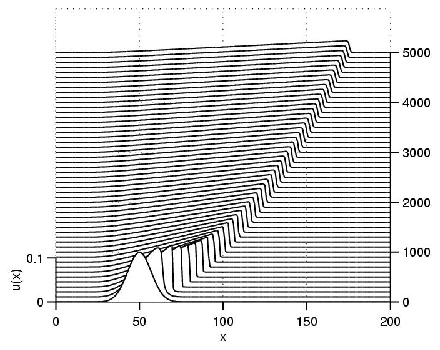, scale=0.7
    }}
}
\caption{\label{Gaussian-to-ramp/cliff-Burgers}
    The Burgers solution is the classic ramp/cliff shown arising from the
    Gaussian initial condition.
}
\end{center}
\end{figure}
}

%\vskip-6.5cm
\remfigure{
\begin{figure}[htbp]
\centering
\psfig{file=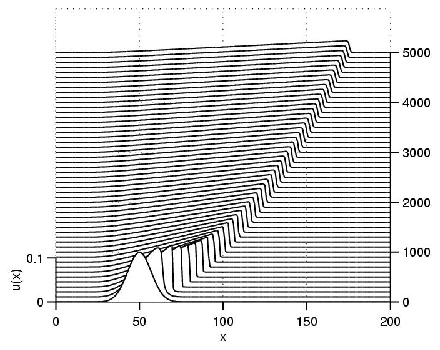,height=6.5cm}
\caption[]{
The Burgers solution is the classic ramp/cliff shown arising from
the Gaussian initial condition.
}
\label{Gaussian-to-ramp/cliff-Burgers}
\end{figure}
}
%%%%%%%%%%%%%%%%%%%%%%%%%%%%%%%%%%%%%%%%%%%%%%%%%%%%%%%%%%%%%%%%%%%%%%%%

The equations KdV, BBM, CH, DP, the other $b-$equations and Burgers all
preserve the area $M=\int_{-\infty}^\infty{u\,dx}$ (total momentum) for a
solution that vanishes at spatial infinity.

\subsection{Problem statement for the present work}
%%%%%%%%%%%%%%%%%%%%%%%%%%%%%%%%%%%%%%%%%%%%%%%%%%%%%%%%%%%%%%%%%%%%%%%%
In this paper, we shall treat the constant $b$ in the dispersionless
case of equation (\ref{b-eqn}) as a {\it bifurcation parameter}. We
shall study numerically the exchange of stability among three types of
solitary wave solutions, as $b$ is varied. This is done by using numerical
simulations to investigate its initial value problem on a periodic domain
of length $L\gg\alpha$ for various values of the parameter $b$. When
necessary, we shall also add viscosity in the form%
%%%%%%%%%%%%%%%%%%%%%%%%%%%%%%%%%%%%%%%%%%%%%%%%%%%%%%%%%%%%%%%%%%%%%%%%
\footnote{{\bf Relation of the $b-$equation (\ref{b-eqn}) to 3D turbulence
closure models.} Higher-dimensional representatives of the viscous
$b-$family of equations (\ref{b-eqn}) have appeared recently in 3D
turbulence modeling. The $b=0$ case of equation (\ref{visc-b-eqn}) is the
1D version of Leray's regularization of the Navier-Stokes equations in 3D
\cite{Leray[1934]}. Leray regularization has recently been revived as an
approach to deriving turbulence closure models for large eddy simulations
of turbulent mixing layers in 3D \cite{GH[2002]}. For 3D studies of the
turbulence model corresponding to the $b-$equation (\ref{b-eqn}) with
$b=2$, see \cite{Chen-etal[1998]},
\cite{Chen-etal[1999c]},
\cite{FHT[2001]}.}
%%%%%%%%%%%%%%%%%%%%%%%%%%%%%%%%%%%%%%%%%%%%%%%%%%%%%%%%%%%%%%%%%%%%%%%%
\begin{equation}\label{visc-b-eqn}
m_t\
+\
\underbrace{\ \ um_x\ \
}
_{\hspace{-2mm}\hbox{convection}\hspace{-2mm}}\
+\
\underbrace{\ \ b\,u_xm\ \
}
_{\hspace{-2mm}\hbox{stretching}\hspace{-2mm}}\
=\
\underbrace{\ \
\nu\,m_{xx}\
}
_{\hspace{-2mm}\hbox{viscosity}
}
\,,
\end{equation}
with $m = u - \alpha^2 u_{xx}$. We shall treat the following cases, which
we shall show analytically are special values:
$b=0,\,\pm1,\,\pm2,\,\pm3$.

\rem{
Recently, the following
equation of this type was singled out for further study based on formal
asymptotics in
\cite{DP[1999]},
\begin{equation}
\label{DP-nonlin}
m_t + um_x + 3mu_x = 0\,,
\hbox{ with }
m = u - \alpha^2 u_{xx}
\,.
\end{equation}
This equation (DP) was shown to be completely integrable by IST and also to
support peakons as soliton solutions in \cite{DHH[2002]}. Subsequently, the
b-family of equations introduced in \cite{DHH[2002]}, \cite{footnote}
\begin{equation}
\label{b-family}
m_t + um_x + bmu_x = 0\,,
\hbox{ with }
m = u - \alpha^2 u_{xx}
\,,
\end{equation}
with constant parameter $b$ was shown \cite{MN[2002]} to admit the
symmetry conditions necessary for integrability only in the cases $b=2$
for CH in (\ref{b-family-inviscid}) and $b=3$ for DP in (\ref{DP-nonlin}).
}

\section{Exchange of stability}
%%%%%%%%%%%%%%%%%%%%%%%%%%%%%%%%%%%%%%%%%%%%%%%%%%%%%%%%%%%%%%%%%%%%%%%%

%\vspace{6.5cm}

%%%%%%%%%%%%%%%%%%%%%%%%%%%%%%%%%%%%%%%%%%%%%%%%%%%%%%%%%%%%%%%%%%%%%%%%
\rem{
\begin{figure}
\begin{center}
   \leavevmode{\hbox{\epsfig{
       figure=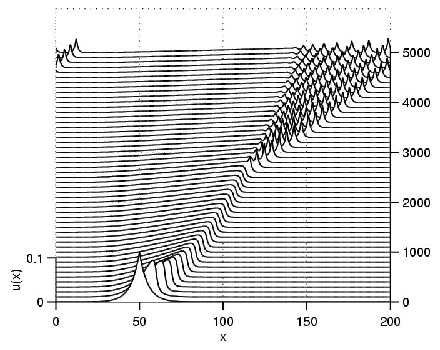, scale=0.7
    }}
}
\caption{\label{peakon-to-ramp/cliff-to-peakons}
    Exchange of stability between the classic ramp/cliff for $b=0$ and the
    peakon for $b=2$.
}
\end{center}
\end{figure}
}

%\vskip-6.5cm
\remfigure{
\begin{figure}[htbp]
\centering
\psfig{file=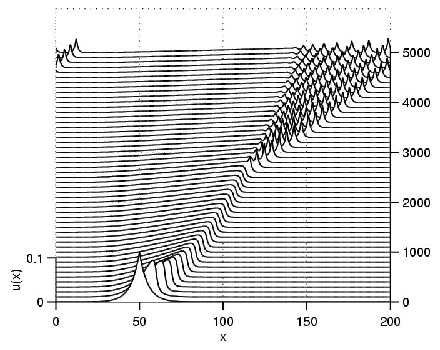,height=6.5cm}
\caption[]{
Exchange of stability between the classic ramp/cliff for $b=0$ and the
peakons for $b=2$.
}
\label{peakon-to-ramp/cliff-to-peakons}
\end{figure}
}
%%%%%%%%%%%%%%%%%%%%%%%%%%%%%%%%%%%%%%%%%%%%%%%%%%%%%%%%%%%%%%%%%%%%%%%%

\subsection{Numerics for exchange of stability from $b=0$ to
$b=\pm2$}
%%%%%%%%%%%%%%%%%%%%%%%%%%%%%%%%%%%%%%%%%%%%%%%%%%%%%%%%%%%%%%%%%%%%%%%%
We begin by considering the initial value problem for case $b=0$ in
equation (\ref{visc-b-eqn}). The first part of Figure
\ref{peakon-to-ramp/cliff-to-peakons} shows for
$b=0$ that a peakon of width $w=5\alpha$ is unstable to
forming a ramp/cliff. Namely, the initial peakon develops into a
Burgers-type ramp/cliff solution for $b=0$, just as it would have done for
the Burgers equation. Note, however, that the case $b=0$ in equation
(\ref{visc-b-eqn}) is {\it not} the Burgers equation. At time $T=2500$, we
stopped this part of the numerical calculation and used the result as the
initial condition for continuing the $b-$equation evolution with $b$ changed
to $b=2$, a case that supports peakons. The remainder of Figure
\ref{peakon-to-ramp/cliff-to-peakons} shows the exchange of stability after
$b=0$ is changed to $b=2$. Thus, the Burgers-type   ramp/cliff solution
is stable for $b=0$ in equation (\ref{visc-b-eqn}). However, when treated
as an initial condition, the ramp/cliff solution is unstable and breaks
into a train of peakons for $b=2$. Figure \ref{final-states/b=0/b=2} shows
the spatial profiles of these solutions at the initial time, at the end of
the
$b=0$ evolution, and at the end of the $b=2$ evolution.  A similar exchange
of stability occurs between the ramp/cliff for $b=0$ and the train of
peakons for $b=3$.

%\vspace{6.5cm}

%%%%%%%%%%%%%%%%%%%%%%%%%%%%%%%%%%%%%%%%%%%%%%%%%%%%%%%%%%%%%%%%%%%%%%%%
\rem{
\begin{figure}
\begin{center}
   \leavevmode{\hbox{\epsfig{
       figure=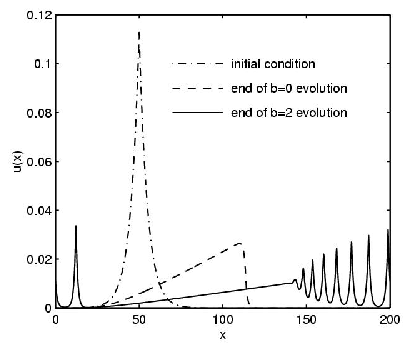, scale=0.7
    }}
}
\caption{\label{final-states/b=0/b=2}
    Spatial profiles of the solutions at the initial time and at the end times
    of the $b=0$ evolution and the $b=2$ evolution.
}
\end{center}
\end{figure}
}

%\vskip-6.5cm
\remfigure{
\begin{figure}[htbp]
\centering
\psfig{file=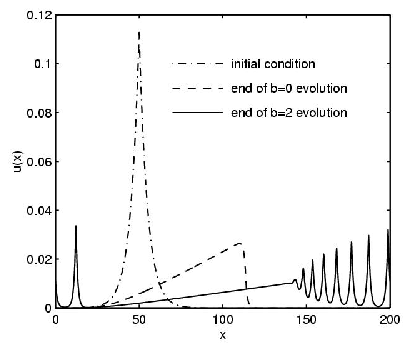,height=6.5cm}
\caption[]{
Spatial profiles of the solutions at the initial time and at the
end times of the $b=0$ evolution and the $b=2$ evolution.
}
\label{final-states/b=0/b=2}
\end{figure}
}
%%%%%%%%%%%%%%%%%%%%%%%%%%%%%%%%%%%%%%%%%%%%%%%%%%%%%%%%%%%%%%%%%%%%%%%%

If we switch from $b=0$ to $b=-2$ instead of $b=2$, then Figure
\ref{peakon-to-ramp/cliff-to-leftons} shows that the ramp/cliff at
time $T=2500$ evolves into a train of {\it leftward moving} solitary waves
(leftons). Thus, a different exchange of stability occurs between
ramp/cliff solutions for $b=0$ and leftons for $b=-2$ (and for
$b=-3$ as well). Figure \ref{final-states/b=0/b=-2} plots the comparison
between
the stationary solution $u=u_0\,{\rm sech}^2\,(x/(2\alpha))$ for $b=-2$
and the final states for an initial Gaussian of width $w=10$ evolved to
leftons.
Similar agreement occurs
for the case $b=-3$, whose stationary solution is $u=u_0\,{\rm
sech}\,(x/\alpha)$. Thus, confined initial states move leftward and evolve
into stationary solutions for $b=-2$ and $b=-3$.

%\vspace{6.5cm}

%%%%%%%%%%%%%%%%%%%%%%%%%%%%%%%%%%%%%%%%%%%%%%%%%%%%%%%%%%%%%%%%%%%%%%%%
\rem{
\begin{figure}
\begin{center}
   \leavevmode{\hbox{\epsfig{
       figure=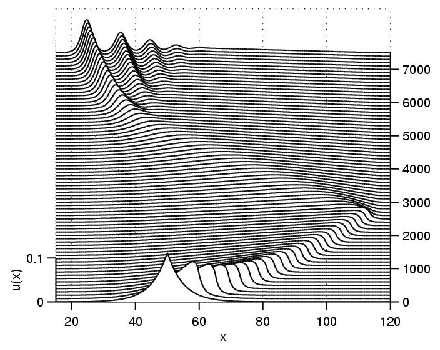, scale=0.7
    }}
}
\caption{\label{peakon-to-ramp/cliff-to-leftons}
    Exchange of stability between the classic ramp/cliff for $b=0$ and the
    leftons for $b=-2$.
}
\end{center}
\end{figure}
}

%\vskip-6.5cm
\remfigure{
\begin{figure}[htbp]
\centering
\psfig{file=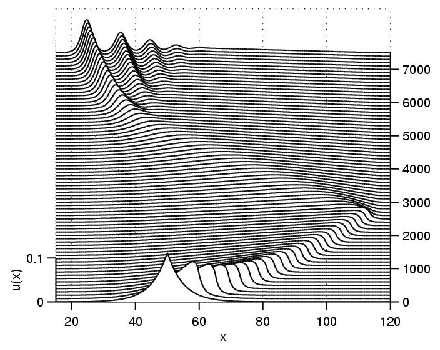,height=6.5cm}
\caption[]{
Exchange of stability between the classic ramp/cliff for $b=0$ and the
leftons for $b=-2$.
}
\label{peakon-to-ramp/cliff-to-leftons}
\end{figure}
}
%%%%%%%%%%%%%%%%%%%%%%%%%%%%%%%%%%%%%%%%%%%%%%%%%%%%%%%%%%%%%%%%%%%%%%%%

%\vspace{6.5cm}

%%%%%%%%%%%%%%%%%%%%%%%%%%%%%%%%%%%%%%%%%%%%%%%%%%%%%%%%%%%%%%%%%%%%%%%%
\rem{
\begin{figure}
\begin{center}
   \leavevmode{\hbox{\epsfig{
       figure=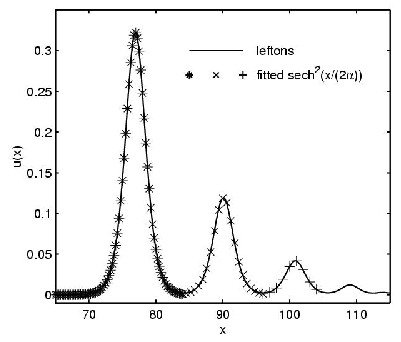, scale=0.7
    }}
}
\caption{\label{final-states/b=0/b=-2}
    Comparison between leftons and the stationary solution
    $u=u_0\,{\rm sech}^2\,(x/(2\alpha))$ for $b=-2$.
}
\end{center}
\end{figure}
}

%\vskip-6.5cm
\remfigure{
\begin{figure}[htbp]
\centering
\psfig{file=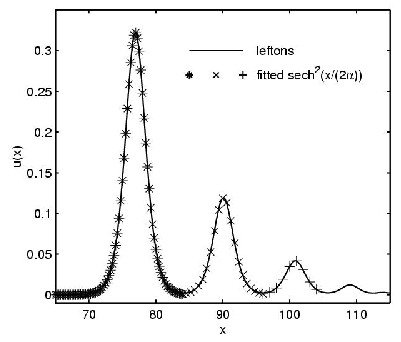,height=6.5cm}
\caption[]{
Comparison between leftons and the stationary solution
$u=u_0\,{\rm sech}^2\,(x/(2\alpha))$ for $b=-2$.
}
\label{final-states/b=0/b=-2}
\end{figure}
}
%%%%%%%%%%%%%%%%%%%%%%%%%%%%%%%%%%%%%%%%%%%%%%%%%%%%%%%%%%%%%%%%%%%%%%%%

\subsection{Analytical discussion of the exchange of stability}
%%%%%%%%%%%%%%%%%%%%%%%%%%%%%%%%%%%%%%%%%%%%%%%%%%%%%%%%%%%%%%%%%%%%%%%%
The exchange of stability as the parameter
$b$ varies can be understood analytically as arising from the following
properties of the $b-$equation (\ref{visc-b-eqn}).

\noindent$\bullet$
For $0\le{b}\le1$, the $L^{1/b}$ norm of the momentum
$\|m\|_{L^{1/b}}=\Big(\int_{-\infty}^\infty|m|^{1/b}\Big)^b$
is preserved by (\ref{visc-b-eqn}), in the inviscid case for which
$\nu=0$. Thus, the ramp/cliff solution for $b=0$ forms in a regime of
classical analytical control in which $|m|$ possesses a maximum principle.
Numerically, we observe: (1) the cliff's width is controlled by the
magnitude of $\alpha$; and (2) the viscosity $\nu$ can be made negligible
without measurably affecting the computed solution.

\noindent$\bullet$
For $1<{b}\le3$ in the inviscid case, an inflection point with negative
slope may be shown to develop into a vertical slope in finite time.
However, numerically, we observe the nonlinear evolution conspires to
eliminate such singular inflection points, e.g., by forming peakons, which
are weak solutions.

\noindent$\bullet$
For $b=-1$ the lowest order nonlinearity in equation (\ref{visc-b-eqn}) (the
steepening term $uu_x$) has coefficient zero. The evolution then develops
from the higher order nonlinear terms. This can be seen by rewriting
the $b-$equation (\ref{visc-b-eqn}) equivalently as
\begin{equation}\label{visc-b-eqn-uform}
u_t + (b+1)uu_x - \nu\,u_{xx}
=
\alpha^2\big(u_t + uu_x - \nu\,u_{xx}+ \frac{b-3}{2}u_x^2\big)_{xx}
\,.
\end{equation}
Thus, solutions with small curvature are nearly stationary for $b=-1$. The
leading order nonlinearity is absent, so the dynamics is governed by the
balance between viscous diffusion and higher-order terms. The case
$\alpha\to\infty$ in (\ref{visc-b-eqn-uform}) is interesting, as well.
However, this case will be discussed elsewhere.

\noindent$\bullet$
Equation (\ref{visc-b-eqn}) has stationary $(c=0)$
solutions $u=u_0\,{\rm sech}^2(x/(2\alpha))$ for $b=-2$ and
$u=u_0\,{\rm sech}(x/\alpha)$ for $b=-3$. Figure
\ref{final-states/b=0/b=-2} shows that the evolution of equation
(\ref{visc-b-eqn}) rapidly approaches the stationary solution for $b=-2$.
(A similar result holds for $b=-3$.) The linearized evolution of
(\ref{visc-b-eqn}) around these stationary solutions is stable for all
wavelengths. Therefore, for any nonzero viscosity, these stationary
solutions have a nontrivial basin of attraction. Note: equation
(\ref{visc-b-eqn}) with either $b=-2$, or $b=-3$ is not Hamiltonian.  The
mathematical properties of the $b-$equation in these cases is discussed
in \cite{HS2003}.

\subsection{Confirmations of our numerical methods}
%%%%%%%%%%%%%%%%%%%%%%%%%%%%%%%%%%%%%%%%%%%%%%%%%%%%%%%%%%%%%%%%%%%%%%%%
In our numerical runs we advanced the $b-$equation (\ref{visc-b-eqn})
with an explicit, variable timestep fourth/fifth order Runge-Kutta-Fehlberg
(RKF45) predictor/corrector.  We selected the timestep for numerical stability
by trial and error, while our code selected the timestep for numerical accuracy
(not to exceed the timestep for numerical stability).  We used a very
strict relative error tolerance per timestep,
$\epsilon=10^{-8}$.

We computed spatial derivatives using 4th order finite differences, generally
at resolutions of $2^{13}=8192$ or $2^{14}=16384$ zones.
To invert the Helmholtz operator in determining $u(x,t)$ from
$m(x,t)$, we used the Fourier transform.  When the numerical approximation
of the nonlinear terms had aliasing errors in the high wavenumbers, we
removed these errors by applying a high pass filtered artificial viscosity.

The quality of the numerical convergence may be checked analytically in the
collision of two peakons of asymptotic speeds $c_1$ and $c_2$, for which the
minimum peakon separation with $b>1$ is
\begin{equation}\label{peakon-qmin}
e^{-|q_{min}|/\alpha} = 1-
\Big(\,\frac{4c_1c_2}{(c_1+c_2)^2}\Big)^{{1}/(b-1)}.
\end{equation}
For $b=2$, $c_1=1$, $c_2=1/2$ and $\alpha=5$, formula (\ref{peakon-qmin})
implies $q_{min}=10\,{\rm ln}\,3=10.9861$.  Our numerical results with the
resolution of $2^{14}$ zones yield $q_{min}=11.0049$.  The small
discrepancy, less than $0.2\%$, occurs largely because our numerical
measurement of $q_{min}$ is obtained by examining the peakon positions at
each internal timestep in the code without interpolation, although one is
unlikely to land exactly on the time at which the minimum separation
occurs.  The code's true accuracy is better than the above measure
indicates, because the intermediate steps involved in advancing the
solution from one discrete time to the next with an RKF45 method cancel the
higher-order discretization errors.

Likewise, for peakons with $b=3$, formula
(\ref{peakon-qmin}) gives the minimum separation when $c_1=1$, $c_2=1/2$,
and $\alpha=5$, as
$q_{min}=5\,{\rm ln}\,(3/(3-\sqrt{8}))=14.3068$. In this case, our numerical
results yield $q_{min}=14.2924$, a discrepancy of only $0.1\%$.

%\vspace{6.5cm}

%%%%%%%%%%%%%%%%%%%%%%%%%%%%%%%%%%%%%%%%%%%%%%%%%%%%%%%%%%%%%%%%%%%%%%%%
\rem{
\begin{figure}
\begin{center}
   \leavevmode{\hbox{\epsfig{
       figure=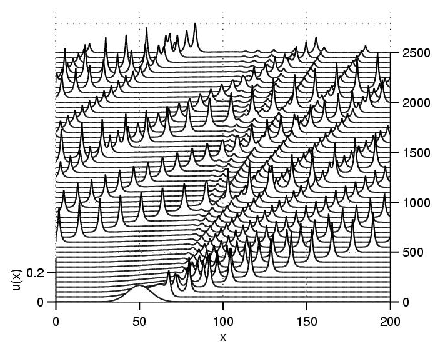, scale=0.7
    }}
}
\caption{\label{scores-of-peakon-collisions/b=3/Gaussian/w=10}
    Scores of peakon collisions arise for (\ref{b-family-inviscid})
    with $b=3$ starting from an initial
    Gaussian velocity distribution of width $w=10$.
}
\end{center}
\end{figure}
}

%\vskip-6.5cm
\remfigure{
\begin{figure}[htbp]
\centering
\psfig{file=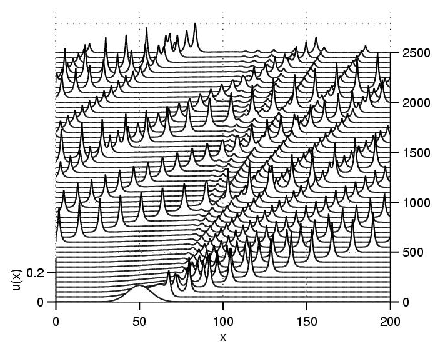,height=6.5cm}
\caption[]{
Scores of peakon collisions arise for $b=3$ (DP) starting from an initial
Gaussian velocity distribution of width $w=10$.
}
\label{scores-of-peakon-collisions/b=3/Gaussian/w=10}
\end{figure}
}
%%%%%%%%%%%%%%%%%%%%%%%%%%%%%%%%%%%%%%%%%%%%%%%%%%%%%%%%%%%%%%%%%%%%%%%%

Of course, the two-peakon collision is rather simple compared to the
plethora of other multi-wave dynamics that occurs in this problem, as in
Figures \ref{scores-of-peakon-collisions/b=3/Gaussian/w=10} and
\ref{DP-peakons-2peakon-ic/w=5/alpha=1}.  For this reason, we also checked
the convergence of our numerical algorithms by verifying that the relative
phases of the peakons in the various figures remained invariant under grid
refinement.  Moreover, the integrity of the waveforms in our figures
attests to the convergence of the numerical algorithm -- after scores of
collisions, the peakon waveforms are still extremely well preserved.  The
preservation of these peakon/soliton waveforms after so many collisions
would not have occurred unless the numerics had converged well.

%\vspace{6.5cm}

%%%%%%%%%%%%%%%%%%%%%%%%%%%%%%%%%%%%%%%%%%%%%%%%%%%%%%%%%%%%%%%%%%%%%%%%
\rem{
\begin{figure}
\begin{center}
   \leavevmode{\hbox{\epsfig{
       figure=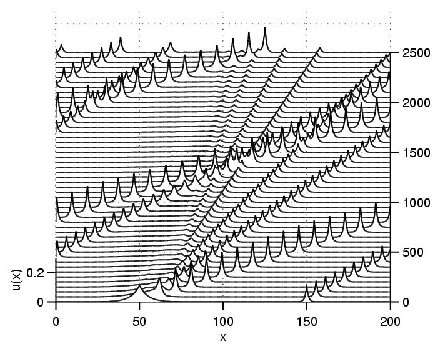, scale=0.7
    }}
}
\caption{\label{DP-peakons-2peakon-ic/w=5/alpha=1}
    Under DP in (\ref{DP-nonlin}) many peakons of unit width for $\alpha=1$
    emerge from  an initial pair of peakons of width $w=5$.
}
\end{center}
\end{figure}
}

%\vskip-6.5cm
\remfigure{
\begin{figure}[htbp]
\centering
\psfig{file=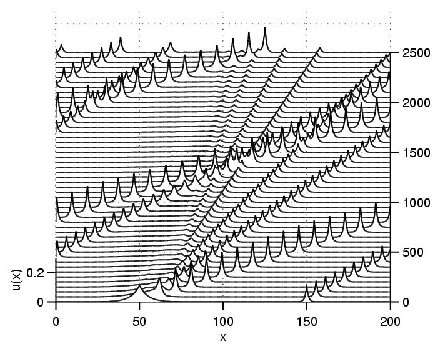,height=6.5cm}
\caption[]{
Under DP in (\ref{b-family-inviscid}) with $b=3$, many peakons of unit
width for $\alpha=1$ emerge from  an initial pair of peakons of width
$w=5$. }
\label{DP-peakons-2peakon-ic/w=5/alpha=1}
\end{figure}
}
%%%%%%%%%%%%%%%%%%%%%%%%%%%%%%%%%%%%%%%%%%%%%%%%%%%%%%%%%%%%%%%%%%%%%%%%

%%%%%%%%%%%%%%%%%%%%%%%%%%%%%%%%%%%%%%%%%%%%%%%%%%%%%%%%%%%%%%%%%%%%%%%%
%%%%%%%%%%%%%%%%%%%%%%%%%%%%%%%%%%%%%%%%%%%%%%%%%%%%%%%%%%%%%%%%%%%%%%%%
\section{Conclusions and further comments}\label{Conclusions}
%%%%%%%%%%%%%%%%%%%%%%%%%%%%%%%%%%%%%%%%%%%%%%%%%%%%%%%%%%%%%%%%%%%%%%%%
%%%%%%%%%%%%%%%%%%%%%%%%%%%%%%%%%%%%%%%%%%%%%%%%%%%%%%%%%%%%%%%%%%%%%%%%

Equation (\ref{b-family-inviscid}) comprises a family of
reversible, parity invariant, evolutionary 1+1 PDEs that arise in the
zero-dispersion case of shallow water waves at quadratic order in the
standard asymptotic expansion,
%%%%%%%%%%%%%%%%%%%%%%%%%%%%%%%%%%%%%%%%%%%%%%%%%%%%%%%%%%%%%%%%%%%%%%%%
\begin{equation}
\label{b-family-inviscid-sum}
m_t + um_x + b\,mu_x = 0\,,
\hbox{ with }
m = u - \alpha^2 u_{xx}
\,.
\end{equation}
%%%%%%%%%%%%%%%%%%%%%%%%%%%%%%%%%%%%%%%%%%%%%%%%%%%%%%%%%%%%%%%%%%%%%%%%
The paper identified the bifurcations and exchanges of stability of its
traveling wave solutions as a function of the nonlinear balance parameter
$b$.

Numerical computations of the initial value problem for
(\ref{b-family-inviscid-sum}) when $b>1$ showed the emergence of its
stable particle-like peakon solutions and their interactions. These may be
obtained analytically by superposing $N$ peakon traveling wave solutions
for the inviscid case,
$u(x,t)=ce^{-|x-ct|}$ as
\begin{equation}\label{peakon-soln-1}
u(x,t)=\sum_{i=1}^Np_i(t)e^{-|x-q_i(t)|}
\quad\hbox{and}\quad
m(x,t)=\sum_{i=1}^Np_i(t)\delta(x-q_i(t))\,,
\end{equation}
for any real constant $b$.  For any $b>1$, the peakons are stable and
undergo particle-like dynamics in terms of the moduli variables $p_i(t)$
and $q_i(t)$, with $i=1,\dots,N$. The peakon dynamics studied for $b>1$
in this framework displayed all of the classical soliton interaction
behavior for peakons found in \cite{CH[1993]}, \cite{FH[2001]} for the case
$b=2$. This behavior included pairwise elastic scattering of peakons,
dominance of the initial value problem by confined pulses and asymptotic
sorting according to height  -- all without requiring complete
integrability.  Thus, the ``emergent pattern'' for $b>1$ in the nonlinear
evolution governed by the $b-$equation (\ref{b-family-inviscid-sum}) was
the rightward moving peakon train, ordered by height.

A second type of emergent pattern of the initial value problem for
(\ref{b-family-inviscid-sum}) occurs in the parameter region
$0\le{b}<1$. This is the classic Burgers-type ramp/cliff structure as in
Figure \ref{final-states/b=0/b=2}. In contrast, for the parameter region
${b}<-1$ a third type of behavior arises, consisting of leftward
moving structures as in Figure \ref{peakon-to-ramp/cliff-to-leftons}.
That the solution behavior should depend on the value of
$b$ is clear from the velocity form of equation
(\ref{b-family-inviscid-sum}) written in (\ref{visc-b-eqn-uform}).

\paragraph{Three regions of $b$.} We found that the solution behavior for
the $b-$equation (\ref{b-family-inviscid-sum})  changes its character near
the two boundaries, $b=\pm1$, of the following three regions in the balance
parameter $b$.
\begin{description}
\item (B1)
In the stable peakon region $b>1$, the Steepening Lemma for the
$b-$equation (\ref{b-family-inviscid-sum}) proven in \cite{CH[1993]} for
$b=2$ and in \cite{HS2003} for $1<b\le3$ allows inflection points with
negative  slopes to escape verticality by producing a jump in spatial
derivative at the peak of a  traveling wave that eliminates the inflection
points altogether. Peakon behavior dominates this region. When $b\le1$ we
found the solution behavior of the $b-$equation
(\ref{b-family-inviscid-sum}) changed its character
and excluded the peakons entirely.

\item (B2)
In the Burgers region $0\le{b}\le1$, the $L^{1/b}$ norm of the
variable $m$ is controlled%
\footnote{For $b=0$, this is a maximum principle for $|m|$.}
and the solution behavior is dominated by ramps and cliffs, as for
the usual Burgers equation.
Similar Burgers-type ramp/cliff solution properties hold for the region
$-1\le{b}\le0$, for which the $L^{1/b}$ norm of the variable $1/|m|$ is
controlled. At the boundary of the latter region, for $b=-1$, the active
transport equation (\ref{b-family-inviscid-sum}) admits stationary plane
waves as exact nonlinear solutions. However, shallow water asymptotic
ordering is broken at $b=-1$.

\item (B3)
In the steady pulse region $b<-1$, pulse trains form that move
leftward from a positive velocity
initial condition (instead of moving rightward, as for $b>-1$). These
leftward-moving pulse trains are found to approach a steady state.

\end{description}

At the two boundaries, $b=\pm1$, of these three regions in $b$, the
emergent solutions of the $b-$family of equations
(\ref{b-family-inviscid-sum}) are observed numerically to exchange
stability. Although we discussed some analytical indications, a full
explanation of these properties remains open for future research.

\subsection*{Acknowledgments}
%%%%%%%%%%%%%%%%%%%%%%%%%%%%%%%%%%%%%%%%%%%%%%%%%%%%%%%%%%%%%%%
We are grateful to A. Degasperis, A.~N.~W.~Hone,
J.~M.~Hyman, S.~Kurien, C.~D.~Levermore, R.~Lowrie and E.~S.~Titi for their
thoughtful insights and remarks. This work was supported by the US DOE under
contracts W-7405-ENG-36 and the Applied Mathematical Sciences Program
KC-07-01-01.

%%%%%%%%%%%%%%%%%%%%%%%%%%%%%%%%%%%%%%%%%%%%%%%%%%%%%%%%%%%%%%%

%\end{multicols}

\end{document}